\documentclass[runningheads]{svmult}

\usepackage{makeidx}   
\usepackage{graphicx}  
\usepackage{subeqnar}  
\usepackage{multicol}  
\usepackage{physprbb}  
\makeindex             

\begin{document}
\title*{Observational tests of the Electro-Magnetic Black Hole Theory in 
Gamma Ray Bursts}
\toctitle{Black holes and GRBs}
%
%
\titlerunning{Observational tests of EMBH theory in GRBs}
%
\author{Remo Ruffini\inst{1}}

\institute{ICRA - International Center for Relativistic Astrophysics and 
Physics Department, University of Rome ``La Sapienza'', I-00185 Rome, 
Italy.} 

\maketitle              

\begin{abstract}
The Relative Space-Time Transformation (RSTT) Paradigm \cite{ruffini_lett1} and the Interpretation of the Burst Structure (IBS) 
Paradigm \cite{ruffini_lett2} are applied to the analysis of the structure of 
the burst and afterglow of Gamma-Ray Bursts within the theory based on the 
vacuum polarization process occurring in an Electro-Magnetic Black Hole, 
the EMBH theory. This framework is applied to the study of the GRB~991216 
which is used as a prototype. The GRB-Supernova Time Sequence (GSTS) 
Paradigm, which introduces the concept of induced gravitational collapse 
in the Supernovae-GRB association \cite{ruffini_lett3}, is announced and will be 
applied, within the EMBH theory, to GRB~980425 as a prototype in a 
forthcoming paper.
\end{abstract}

\section{Introduction}

I am very pleased to present here in Munich some observational tests of 
our Electro-Magnetic-Black-Hole theory, for short the EMBH theory, 
explaining some features of Gamma Ray Bursts, for short GRBs. The EMBH 
theory is rooted in discussions I had from 1971 to 1975 with Werner 
Heisenberg here in Munich, in Washington and Stanford. 

GRBs are today promoting one of the most ample scientific effort in the 
entire field of science, both in the observational and theoretical 
domains. Following the discovery of the GRBs by the Vela satellites 
\cite{ruffini_s75}, the observations from the Compton satellite and BATSE had 
shown the isotropical distribution of the GRBs strongly suggesting a 
cosmological nature for their origin. It was still trough the data of 
BATSE that the existence of two families of bursts, the ``short bursts'' 
and the ``long bursts'' was presented, opening an intense scientific 
dialogue on their origin still active today, as we see in the talk 
of M. Schmidt in these proceedings.

\begin{figure}
\begin{center}
\includegraphics[width=12cm,clip]{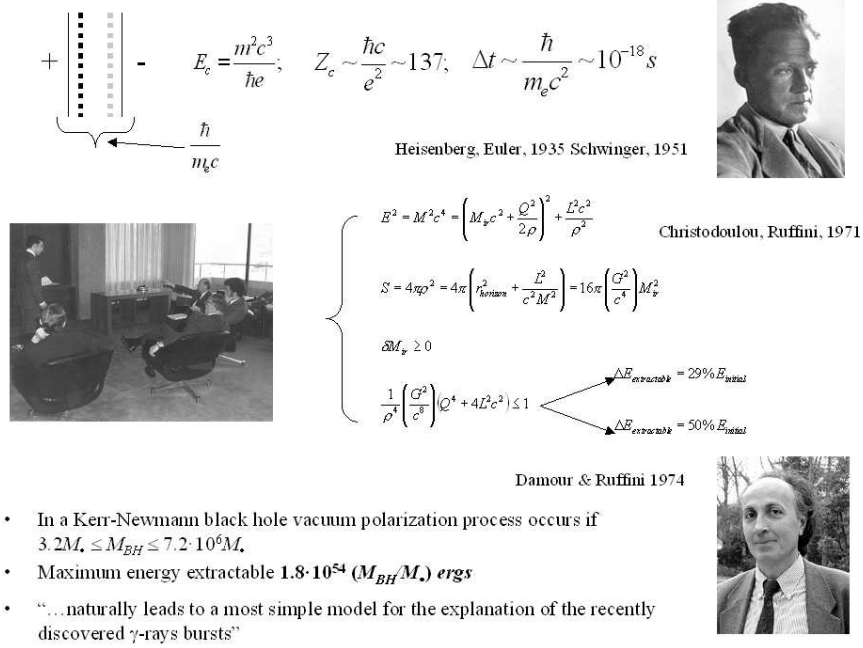} 
\end{center}
\caption{Theoretical background of the EMBH model. The critical field against the breakdown of the vacuum is given next to a picture of Werner Heisenberg. The mass formula of EMBH is given next to the picture of the Ph.D. thesis discussion in Princeton by Demetrios Christodoulou, at the age of 19, in front of the committee with members: R. Ruffini (supervisor), J.A. Wheeler, E. Wigner and D. Wilkinson. The energetics of the vacuum polarization process of an EMBH is given next to the picture of T. Damour. These results were obtained, with R. Ruffini as supervisor, at Princeton University in 1974 during the preparation of the thesis of Doctorat d' Etat finally presented in Paris.}
\label{ruffini_3theories} 
\end{figure}

An enormous momentum was gained in this field by the discovery of the 
afterglow phenomena by the BeppoSAX satellite and the optical 
identification of the GRBs sources at cosmological distances (see e.g. 
\cite{ruffini_c00}). It has become apparent that fluxes of $10^{54}$ ergs/s are 
reached: during its peak emission the energy of a single GRB equals the 
energy emitted by all the stars of the Universe (see e.g. \cite{ruffini_rk01}).

From an observational point of view, an unprecedented campaign of 
observations is at work using the largest deployment of observational 
techniques from space with the satellites CGRO - BATSE, Beppo-SAX, Chandra, 
R-XTE, XMM-Newton, HETE-2, as well as the HST, and from the ground with 
optical KECK and the VLT and radio  by the VLA observatories. Possibility 
of further examining correlation with detection of ultrahigh energy 
cosmic rays and neutrinos should be reachable in the near future thanks to 
developments of the AUGER and AMANDA experiments (see also \cite{ruffini_h00}).

\begin{figure}
\begin{center}
\includegraphics[width=12cm,clip]{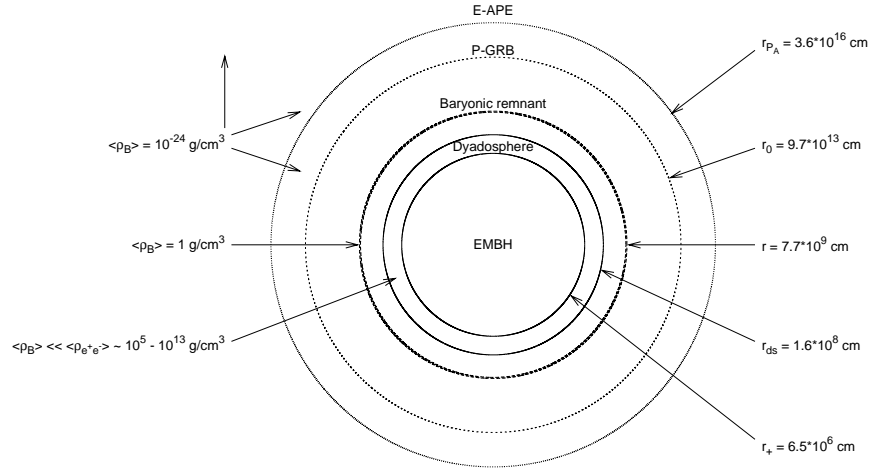}
\end{center}
\caption{Selected events of the expansion of the pulse in the EMBH model 
are represented, together with the distance from the EMBH in the 
laboratory frame at which they occur and with the density of the 
surrounding medium. The radial distances are represented in logarithmic 
scale.}
\label{ruffini_raggi2}
\end{figure}

From a theoretical point of view, the GRBs offers comparable 
opportunities to develop new domains in yet untested directions of 
fundamental science. For the first time within the EMBH model, see 
Fig.~\ref{ruffini_raggi2}, the opportunity exist to 
theoretically approach the following fundamental issues:
\begin{enumerate}
\item extreme relativistic hydrodynamic phenomena of an electron positron 
plasma expanding with sharply varying Lorentz gamma factors in the range 
$10^2$ to $10^4$. Analyze as well the very high collision of such 
expanding plasma with baryonic matter reaching intensities $10^{40}$ larger 
then the ones usually obtained in Earth based accelerators, see \cite{ruffini_rbcfx02} and references therein,
\item the bulk process of vacuum polarization created by overcritical 
electromagnetic fields, in the sense of Heisenberg, Euler \cite{ruffini_he35} 
and Schwinger \cite{ruffini_s51}. This long sought quantum ultrarelativistic 
effect, not yet convincingly observed in heavy ion collision on the Earth 
in Darmstadt, may indeed make its first appearance in the strong 
electromagnetic fields developed in astrophysical conditions during the 
process of gravitational collapse to a black hole, see \cite{ruffini_prx98,ruffini_rswx99,ruffini_rswx00},
\item a novel form of energy source: the extractable energy of a black 
hole, introduced by Christodoulou and Ruffini \cite{ruffini_cr71}.  The enormous 
energies, released almost instantly in the observed GRBs, points to the 
possibility that for the first time we are witnessing the release of the 
extractable energy of an EMBH, during the process of gravitational 
collapse itself. We can compute and, if observationally confirmed, have 
the opportunity to study all general relativistic effects as the horizon 
of the Black hole is approached and is being formed, together with the 
associated ultrahigh energy quantum phenomena, see \cite{ruffini_crv02,ruffini_rv02,ruffini_rcvx02}.
\end{enumerate}

It is clear that in the approach to such a vast new field of research 
implying previously unobserved relativistic regimes it is not possible to 
proceed {\em as usual} adopting an uncritical comparison of observational data to 
theoretical models within the classical schemes of astronomy and 
astrophysics. Some insight to the new approach needed can be gained from 
past experiences in the interpretation of relativistic effects in high 
energy particle physics as well as from the explanation of some 
relativistic effects observed in the astrophysical domain. All those 
relativistic regimes are however much less extreme then the new ones encountered in 
GRBs.

There are at least three major new features in relativistic systems which have to 
be taken into due account:

\begin{figure}
\begin{center}
\includegraphics[width=10cm,clip]{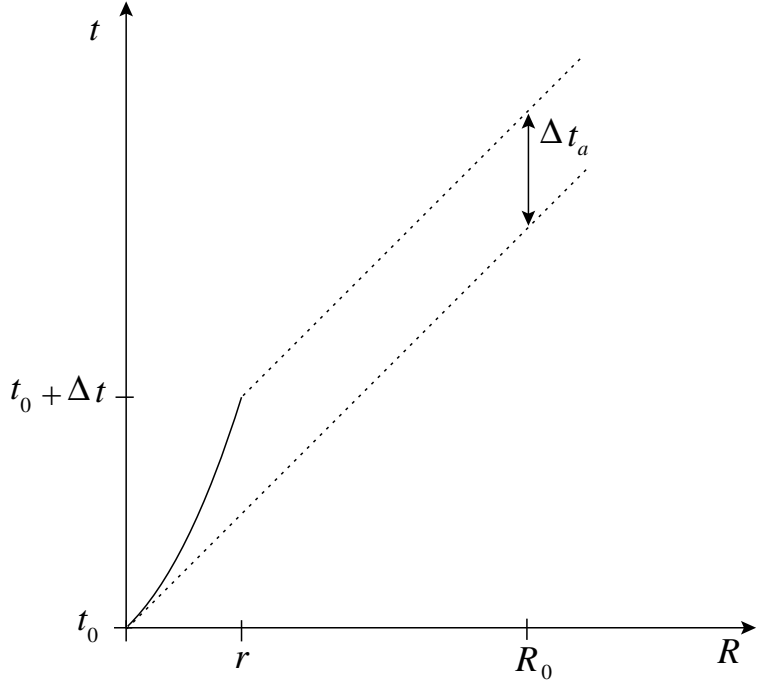}
\end{center}
\caption{This qualitative diagram illustrates the relation between the 
laboratory time interval $\Delta t$ and the arrival time interval $\Delta 
t_a$ for a pulse moving with velocity $v$ in the laboratory time (solid 
line). We have indicated here the case where the motion of the source has 
a nonzero acceleration. The arrival time is measured using light signals 
emitted by the pulse (dotted lines). $R_0$ is the distance of the observer 
from the EMBH, $t_0$ is the laboratory time corresponding to the onset of 
the gravitational collapse, and $r$ is the radius of the expanding pulse 
at a time $t=t_0 + \Delta t$. See also \cite{ruffini_lett1}.}
\label{ruffini_ttasch}
\end{figure}

\begin{enumerate}
\item Practically the totally of data in astronomical and astrophysical 
systems are acquired by using photons arrival time. It was Einstein 
\cite{ruffini_e05} at the very initial steps of special relativity to caution on 
the use of such an arrival time analysis and to state that, when dealing 
with objects in motion, due care should be put in defining time 
synchronization procedure in order to construct the correct spacetime 
coordinate grid (see Fig.~\ref{ruffini_ttasch}). It is not surprising that as soon 
as the first bulk motions relativistic effects were observed by radio and optical telescopes their 
interpretations within the classical framework of astrophysics led to the 
concept of ``superluminal'' motion. These observations refer to 
extragalactic radio sources, with Lorentz gamma factors $\sim 6$ 
\cite{ruffini_bsm99} and to microquasars in our own galaxy with Lorentz gamma 
factor $\sim 5$ \cite{ruffini_mr99}. It has been recognized \cite{ruffini_r66} that no 
``superluminal'' motion exist as soon as the prescriptions indicated by 
Einstein are used in order to establish the correct spacetime grid for the 
astrophysics systems. In the present context of GRBs, where the Lorentz 
gamma factor can easily surpass $10^2$, the direct application of 
classical concepts leads to enormous ``superluminal'' behaviors 
\cite{ruffini_rbcfx02}. An approach based on classical arrival time considerations, 
as done sometime in current literature, completely subvert the causal 
relation in the observed astrophysical phenomenon.
\item One of the clear success of relativistic field theorists has been 
the understandings of the role of four momentum energy conservation laws 
in multiparticle collisions and decays such as the reaction: 
$n\rightarrow p+e+\bar\nu$. From the works of Pauli anf Fermi it became clear 
how in such processes, contrary to the case of classical mechanics, it is 
impossible to analyze a single term of the decay, the electron or the 
proton or the neutrino or the neutron, out of the context of the global 
point of view of the relativistic conservation of the total four momentum 
of the system, which involves the knowledge of the system during the 
entire decay process. These rules are routinely used by workers in high 
energy particle physics and have become part of their cultural 
background. If we apply these same rules to the case of the relativistic 
system of a GRB it is clear that it is just impossible to consider a part 
of the system, e.g. the afterglow, out of the general conservation laws and 
history of the {\em entire relativistic regime} of the system. The description of the sole afterglow, as 
has been done at times in the literature, could indeed be done within the 
framework of classical astronomy and astrophysics, but not in a 
relativistic astrophysics where the entire space-time grid necessary for 
the description of the afterglow depends on all the previous relativistic 
part of the worldline of the system.
\item The very lifetime of a phenomenon has not an absolute meaning, 
special and general relativity have shown. It depends both from inertial 
reference frame of the laboratory and of the observer and their relative 
motion. Such a phenomenon, generally expressed in the ``twin paradox'' has 
been extensively checked and confirmed to extreme high accuracy in 
elementary particle physics in the CERN experiments. This situation is 
much more extreme in GRBs due to the very large (in the range $10^2-10^4$) 
and time varying (on time scales ranging from fractions of seconds to 
months) gamma Lorentz factors between the Laboratory frame and the far 
away observer. Such an observer is moreover in the GRBs context further 
affected by the cosmological recession velocities of its local Lorentz 
frame.
\end{enumerate}

\begin{figure}
\begin{center}
\includegraphics[width=12cm,clip]{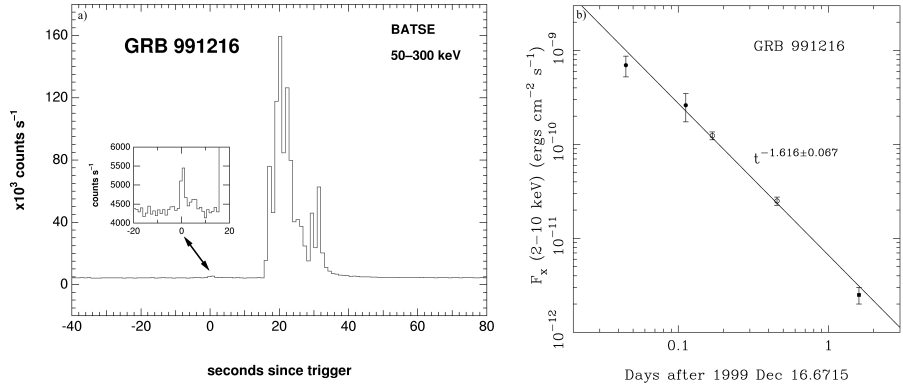}
\end{center}
\caption{{\bf a)} The peak emission of GRB~991216 as seen by BATSE 
(reproduced from \cite{ruffini_brbr99}); {\bf b)} The afterglow emission of 
GRB~991216 as seen by XTE and Chandra (reproduced from \cite{ruffini_ha00}).}
\label{ruffini_grb991216}
\end{figure}

These are some of the reasons why we have recently presented a basic 
Relative Space-Time Transformations RSTT paradigm \cite{ruffini_lett1} to be 
applied prior to the interpretation of GRBs data. 

The first step is the establishment of the constitutive equations 
relating:\\
a) The comoving time of the pulse ($\tau$)\\
b) The laboratory time ($t$)\\
c) The arrival time ($t_a$)\\
d) The arrival time at the detector ($t_a^d$)\\
The book-keeping of the four different times and corresponding space 
variables must be done carefully in order to keep the correct causal 
relation in the time sequence of the events involved. 

The RSST paradigm reads: {\em ``the necessary condition in order to 
interpret the GRB data is the knowledge of the entire worldline of the 
source from the moment of gravitational collapse. In order to meet this 
condition, given a proper theoretical description and the correct 
constitutive equations, it is sufficient to know the energy of the 
dyadosphere and the mass of the remnant of the progenitor star''}.

Clearly such an approach is at variance with the works in the current 
literature which mainly emphasize either some qualitative description of 
the sources or some quantitative phenomenological multiparameter fitting 
of the sole afterglow era. 

\begin{itemize}
\item Many works in the literature have addressed the issue of the 
sources of GRBs. They include works on binary neutron stars mergers (see 
e.g. \cite{ruffini_elps89,ruffini_npp92,ruffini_mr92mnras,ruffini_mr92apj}), black hole - white dwarf 
\cite{ruffini_fwhd99} and black hole - neutron star binaries \cite{ruffini_p91,ruffini_mr97b}, 
Hypernovae (see \cite{ruffini_p98}), failed supernovae or collapsars (see 
\cite{ruffini_w93,ruffini_mw99}), supranovae (see \cite{ruffini_vs98,ruffini_vs99}). Only those based on 
binary neutron stars have reached the definition of detailed quantitative 
estimates of a model, but they present serious difficulties in the energetics, as 
well as in the explanation of ``long bursts'' (see 
\cite{ruffini_swm00,ruffini_wmm96}), and in the observed location of the GRBs' sources in 
star forming regions (see \cite{ruffini_bkd00}). In the remaining cases was 
presented a sole qualitative analysis of the sources without addressing the 
overall problem from the source to the observations: the necessary details 
to formulate the equations of the dynamical evolution of the system are 
generally still missing.
\end{itemize}

Other works in the literature have mainly addressed the problem of 
fitting the data of the afterglow observations by phenomenological 
analysis. They are separated in two major classes:
 
\begin{itemize}
\item The ``internal shock models'', first introduced by \cite{ruffini_rm94}, are by 
far the most popular ones having been developed in many different aspects, 
e.g. by \cite{ruffini_px94,ruffini_sp97,ruffini_f99,ruffini_fcrsyn99}. The underlying assumption is that 
all the variabilities of GRBs in the range $\Delta t\sim 1$ 
ms up to the overall duration $T$ of the order of $50$ s are determined by 
a yet undetermined ``inner engine''. The difficulties of explaining the 
long time scale bursts by a single explosive event has evolved into a 
variety of assumptions on yet unspecified family of``inner engines'' with a prolonged activity 
(see e.g. \cite{ruffini_p01} and references therein).
\item The ``external shock models'', also introduced by \cite{ruffini_mr93}, are less popular today. There is the distinct possibility, within these models, 
that ``GRBs' light curves are tomographic images of the density 
distribution of the medium surrounding the sources of GRBs'' (\cite{ruffini_dm99}, 
see also \cite{ruffini_dcb99,ruffini_d00} and references therein). It is generally outlined that the 
structure of the burst does not depend directly from the ``inner engine''.
\end{itemize}

\begin{figure}
\begin{center}
\includegraphics[width=10cm,clip]{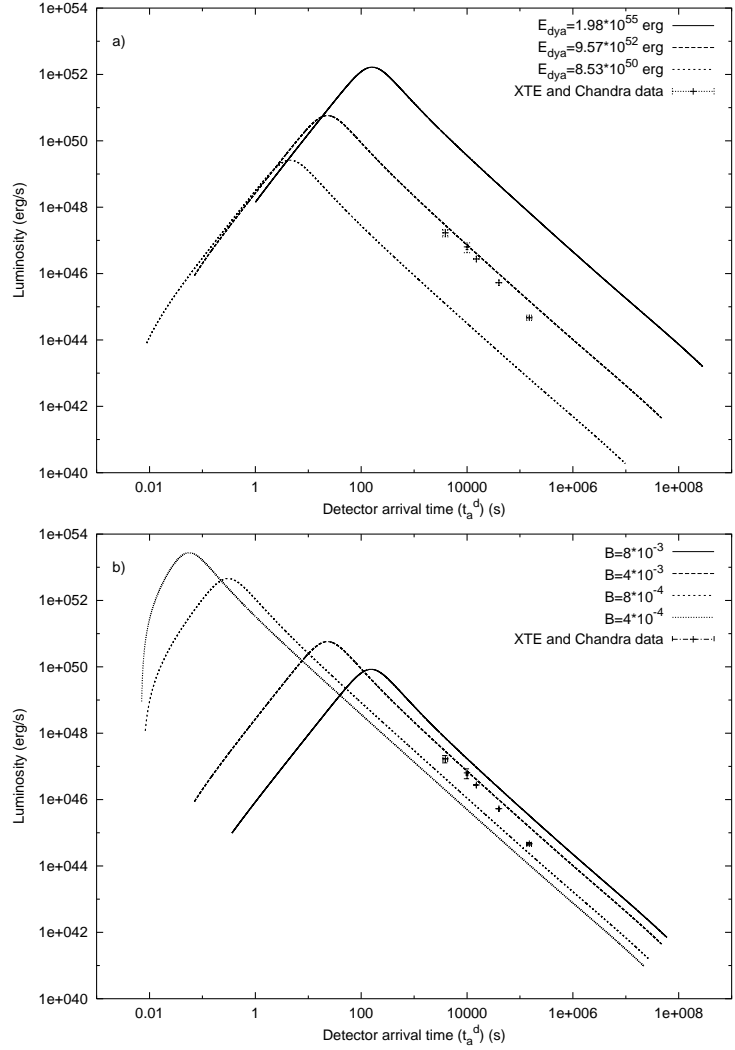}
\end{center}
\caption[]{a) Afterglow luminosity computed for an EMBH of $E_{dya}= 
8.53\times 10^{50}$ ergs, $E_{dya}= 9.57\times 10^{52}$ ergs, $E_{dya}= 
1.98 \times 10^{55}$ ergs and $B=4\times 10^{-3}$. b) for the $E_{dya}= 
9.57\times 10^{52}$, we give the afterglow luminosities corresponding 
respectively to $B=4\times 10^{-4}$, $8\times 10^{-4}$, $4\times 10^{-3}$, 
$8\times 10^{-3}$.}
\label{ruffini_ii-fig2}
\end{figure}

All these works encounter the above mentioned difficulties, they present 
only a piecewise description of the GRB phenomenon and by neglecting the 
earlier phases, their space time grid is undefined and, as we have shown 
in detail in \cite{ruffini_rbcfx02}, results are reached at variance from the 
ones obtained in a complete and unitary description of the GRBs 
phenomenon. We outline in the following how such an unitary description 
naturally leads to new characteristic features both in the burst and 
afterglow of GRBs.

In a series of papers, we have developed the above mentioned EMBH model 
\cite{ruffini_rukyoto} which has the great advantage, despite its simplicity, 
that all eras of the model, following the process of gravitational 
collapse, are described by precise field equations which can then be 
numerically integrated. The three basic starting points are:
\begin{itemize}
\item the extractable energy of an EMBH introduced in \cite{ruffini_cr71},
\item the vacuum polarization process {\it \`a la} Heisenberg-Euler-Schwinger \cite{ruffini_he35,ruffini_s51} 
in the field of an EMBH, first computed in \cite{ruffini_dr75},
\item the fact that vacuum polarization process can indeed be a realization of the 
reversible transformation of an EMBH introduced in \cite{ruffini_cr71}.
\end{itemize}
These were the themes of discussions with Werner Heisenberg. He was supposed to 
inaugurate the 1975 Varenna Summer School \cite{ruffini_gr75} directed to these 
subjects if he had not died a few weeks earlier. In that school and in 
\cite{ruffini_dr75} the possibility that the process in \cite{ruffini_he35,ruffini_s51} 
duly extended to the EMBH were at the very basis of the 
explanation of GRBs was advanced (see 
Fig.~\ref{ruffini_3theories}).

Following the Beppo SAX observations and the energetics requirements, we have returned to our EMBH model \cite{ruffini_rukyoto} and developed  the dyadosphere 
concept \cite{ruffini_prx98}. The dynamics of the $e^+e^-$-pairs and 
electromagnetic radiation of the plasma generated in the dyadosphere 
propagating away from the EMBH in a sharp Pairs-Electro-Magnetic pulse 
(the PEM pulse) has been studied by us and validated by the numerical 
codes at Livermore Lab \cite{ruffini_rswx99}. The collision of the still 
optically thick PEM pulse with the baryonic matter of the remnant of the 
progenitor star has been again studied and validated by the Livermore Lab 
codes \cite{ruffini_rswx00}. The further evolution of the sharp pulse of 
pairs-electromagnetic radiation and baryons (the PEMB pulse) further 
proceeds with increasing values of the Lorentz gamma factor until the 
condition of transparency is reached \cite{ruffini_brx00}. At this stage the 
Proper-Gamma Ray Burst (P-GRB) is emitted \cite{ruffini_lett2} and a pulse of 
Accelerated-Baryonic-Matter the (ABM pulse) is injected in the 
interstellar medium giving rise to the afterglow.

The interaction of the ABM-Pulse giving origin to the afterglow has been 
recently developed and presented in detail in \cite{ruffini_rbcfx02}. We recall  
the minimum set of assumptions we have adopted:

\begin{enumerate}
\item the collision of the ABM pulse is assumed to occur with a constant 
homogeneous interstellar medium of number density $n_{\rm ism} \sim 1 
{\mathrm cm}^{-3}$. The energy emitted in the collision is assumed to be 
instantaneously radiated away (fully radiative condition). The description 
of the collision and emission process is done in spherical symmetry, 
taking only the radial approximation neglecting all the delay emission by 
scattered radiation.
\item special attention is given to numerically compute the power of the 
afterglow as a function of the arrival time using the correct constitutive 
equations for the space-time transformations in line with the RSTT 
paradigm.
\item finally some approximate solutions are adopted in order to obtain 
the determination of the power law indexes of the afterglow flux and 
compare and contrast them with the observational results as well as with 
the alternative results in the literature. 
\end{enumerate}
In \cite{ruffini_rbcfx02} we have considered uniquely the above radial 
approximation in order to concentrate on the special role of the correct 
space time transformations in the RSST paradigm and to explicitly 
illustrate their impact on the determination of the power law index of the 
afterglow. This topic has been unduly neglected in the literature. We 
enter in a forthcoming papers both in the details of the role of beaming 
of the radiation and of the diffusion due to off axis emission 
\cite{ruffini_lettS,ruffini_bcrx02}.

\begin{figure}
\begin{center}
\includegraphics[width=10cm,clip]{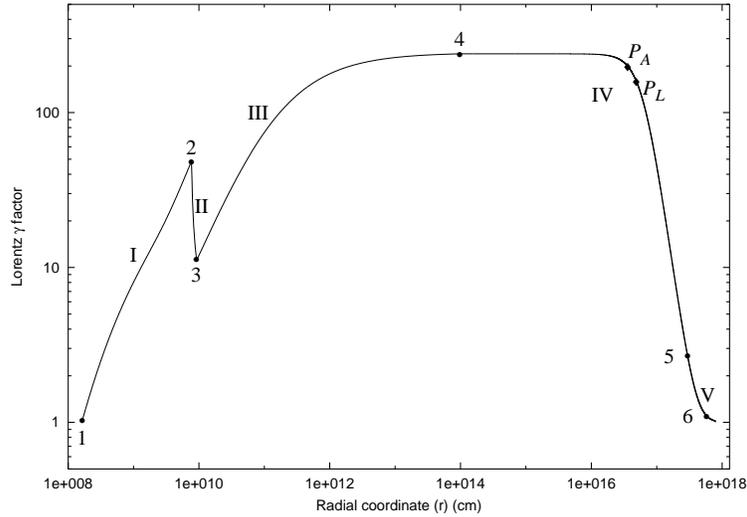}
\end{center}
\caption{The theoretically computed Lorentz gamma factor for the parameter 
values $E_{dya}=9.57\times 10^{52}$ erg, $B=4\times 10^{-3}$ is given as a 
function of the radial coordinate in the laboratory frame. The 
corresponding values in the comoving time, laboratory time and arrival 
time are given in \cite{ruffini_rbcfx02}. The different eras, indicated by roman 
numerals, are illustrated in the text, while the points 1,2,3,4,5 mark the 
beginning and end of each of these eras. The points $P_L$ and $P_A$ mark 
the maximum of the afterglow flux, respectively in emission time and in 
arrival time (see \cite{ruffini_lett2}). The point 6 is the beginning of Phase D 
in era V. At point 4 the transparency condition is reached.}
\label{ruffini_gamma}
\end{figure}

It is now clear after the observations of GRB~980425 that the afterglow 
phenomena can present, especially in the optical and radio wavelengths, 
features originating from phenomena spatially and causally distinct from 
the GRB phenomena. There is evidence in four different GRB systems, including GRB~991216 due to the observed emission in the iron lines and their shift, that a second component exist associated to the GRBs: a supernova. This supernova explosion follows in time the GRB emission and occurs at distances of $10^{16}-10^{17}$ cm away from the location where the gravitotional collapse to the EMBH has occurred. The space time analysis of these events can be correctly performed when the RSTT paradigm is adopted \cite{ruffini_lett1,ruffini_lett2}. These facts have motivated us to introduce the novel concept of induced 
supernovae \cite{ruffini_lett3}. In the current litterature the existence of these two different components has not been recognized and attempts have been made of fitting the data of the supernova, in the x-rays as well as in the optical and radio, as part of the afterglow within the framework of a multiparameters fitt. For the above mentioned reasons, such an approach adds further difficulties to the already critical situation of the current literature.

We have therefore confronted the theoretical 
results of the EMBH model with the data of GRB~991216 as a prototypical 
case (see Fig.~\ref{ruffini_grb991216}). The reason of this choice are simply 
given:
\begin{enumerate}
\item This is one of the strongest GRBs in x-rays and is also quite 
general in the sense that shows relevant cosmological effects. It radiates 
mainly in X-rays and in $\gamma$-rays and less then 3\% is emitted in 
optical and radio bands (see \cite{ruffini_ha00}). Also the emission of the supernova, inferred from the iron lines, is in this case  weaker then the autentic GRB energy flux.
\item The excellent data obtained by BATSE on the burst \cite{ruffini_brbr99} 
are complemented by the data on the afterglow acquired by the Chandra 
\cite{ruffini_p00} and RXTE \cite{ruffini_cs00}, and also superb data have been obtained from 
spectroscopy of the iron lines \cite{ruffini_p00}.
\item A very precise value for the slope of the energy emission during the 
afterglow as a function of time has been obtained: $n=-1.64$ 
\cite{ruffini_tmmgk99} and $n=-1.616\pm 0.067$ \cite{ruffini_ha00}.
\end{enumerate}

The comparison of the EMBH model to the data of the GRB~991216 and its 
afterglow has naturally led to a new paradigm for the interpretation of 
the burst structures (IBS paradigm) of GRBs \cite{ruffini_lett2}. The IBS 
paradigm reads: {\em ``in GRBs we can distinguish an injector phase and a 
beam-target phase. The injector phase includes the process of 
gravitational collapse, the formation of the dyadosphere, as well as era I 
(the PEM pulse), era II (the engulfment of the baryonic matter of the 
remnant) and era III (the PEMB pulse). The injector phase terminates with 
the P-GRB emission. The beam-target phase addresses the interaction of the 
ABM pulse, namely the beam generated during the injection phase, with the 
ISM as the target. It gives rise to the E-APE and the decaying part of the 
afterglow''}.

\begin{figure}
\begin{center}
\includegraphics[width=10cm,clip]{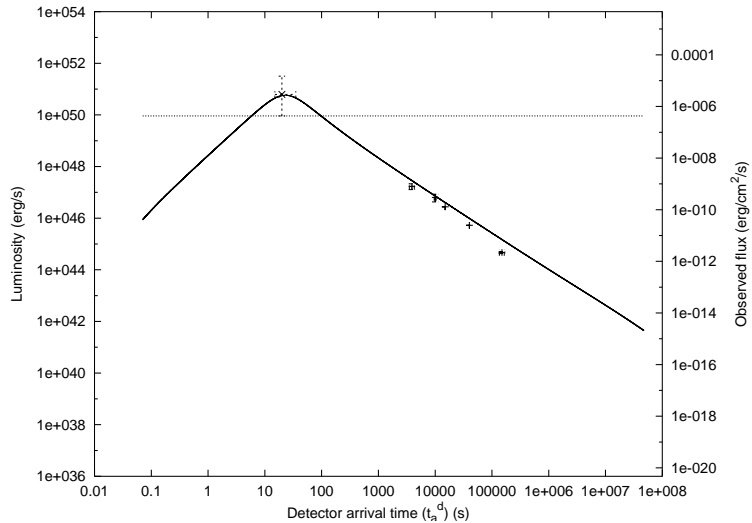}
\end{center}
\caption{Best fit of the afterglow data of Chandra, RXTE as well as of the 
range of variability of the BATSE data on the major burst, by a unique 
afterglow curve leading to the parameter values $E_{dya}=9.57\times 
10^{52}erg, B=4\times 10^{-3}$. The horizontal dotted line indicates the 
BATSE noise threshold. On the left axis the luminosity is given in units 
of the energy emitted at the source, while the right axis gives the flux 
as received by the detectors.}
\label{ruffini_fit_1}
\end{figure}

We recall that:\\
{\bf a)} The {\bf injector phase} starts from the moment of gravitational 
collapse all the way to the emission of the proper GRB (the P-GRB) and 
encompasses the following eras:
\begin{itemize}
\item {\em The zeroth era}: the formation of the dyadosphere;
\item {\em The era I}: the expansion of the PEM pulse; 
\item {\em The era II}: the interaction of the PEM pulse with the remnant 
left over by the collapse of the progenitor star;
\item {\em The era III}: the further expansion of the PEMB pulse; 
The injector phase is concluded by the emission of the P-GRB and the 
ABM pulse, as the condition of transparency is reached.
\end{itemize}
{\bf b)} The {\bf beam-target phase}, in which the accelerated baryonic 
matter (ABM) generated in the injector phase collides with the 
interstellar medium (ISM), gives origin to the afterglow and encompasses 
the following eras:
\begin{itemize}
\item {\em The Era IV}: the ultra relativistic and relativistic regimes in 
the afterglow: the emitted flux first increases to reach a maximum value 
and then monotonically decrease following well defined power laws in the 
arrival time;
\item {\em The Era V}: the approach to the non relativistic regimes in the 
afterglow, also describable by specific power laws in the arrival time;
\end{itemize}
Some qualitative representation of these eras as a function of the radial 
coordinate, in logarithmic scale are represented in Fig.~\ref{ruffini_raggi2}.

\begin{figure}
\begin{center}
\includegraphics[width=10cm,clip]{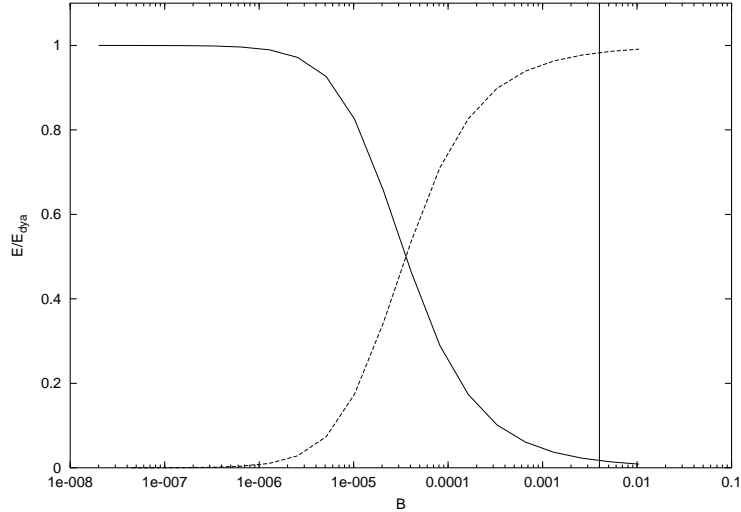}
\end{center}
\caption{Relative intensities of the E-APE (dashed line) and the P-GRB 
(solid line), as predicted by the EMBH model corresponding to the values 
of the parameters determined in Fig.~\ref{ruffini_fit_1}, as a function of $B$. 
Details are given in \cite{ruffini_rbcfx02}. The vertical line corresponds to the 
value $B=4\times 10^{-3}$.}
\label{ruffini_crossen}
\end{figure}

The comparison of the EMBH theory to the data of GRB~991216 has allowed 
the determination of the only two free parameters of the theory: the energy 
of the Dyadosphere $E_{dya}$ and the mass of the baryonic remnant left over 
by the collapse of the progenitor star, measured in units of the 
$E_{dya}$, defined by the dimensionless $B$ parameter. Details are given in \cite{ruffini_rbcfx02}.

We have then obtained, for the first time, the complete history 
of the Lorentz gamma factor from the moment of gravitational collapse to 
the latest phases of the afterglow observations (see Fig.~\ref{ruffini_gamma}). We 
have as well determined the entire space time grid of the GRB~991216 by 
giving (see Table~1 in \cite{ruffini_rbcfx02}) the radial coordinate of the GRBs 
phenomenon as a function of the four coordinate time variables. The 
extreme relativistic regimes at work in GRB~991216 lead to enormous 
superluminal behavior (up to $10^4 c$!) if the classical astrophysical 
concepts were adopted using the arrival time as the independent variable (see Table~1 in \cite{ruffini_rbcfx02}). 
In turn this implies that any causal relation based on classical 
astrophysics and the arrival time data, as often done in current GRBs 
literature, is incorrect.

\begin{figure}
\begin{center}
\includegraphics[width=10cm,clip]{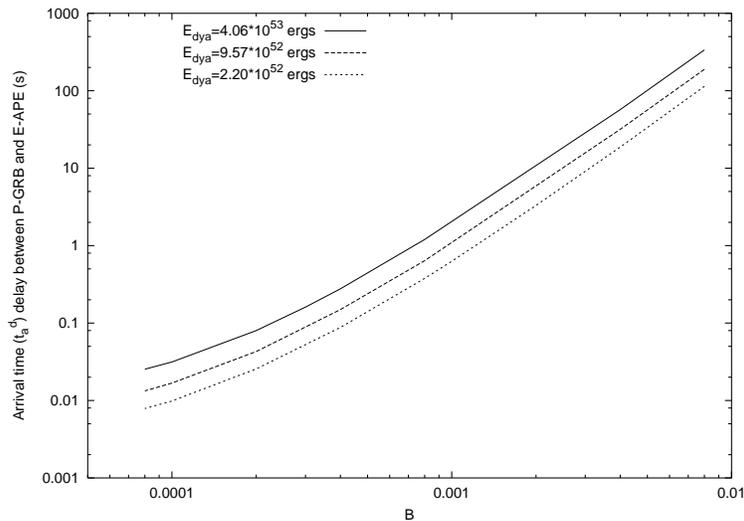}
\end{center}
\caption{Time delays between P-GRB and E-APE as a function of the B 
parameter for selected values of the $E_{dya}$.}
\label{ruffini_dtab}
\end{figure}

\begin{figure}
\begin{center}
\includegraphics[width=10cm,clip]{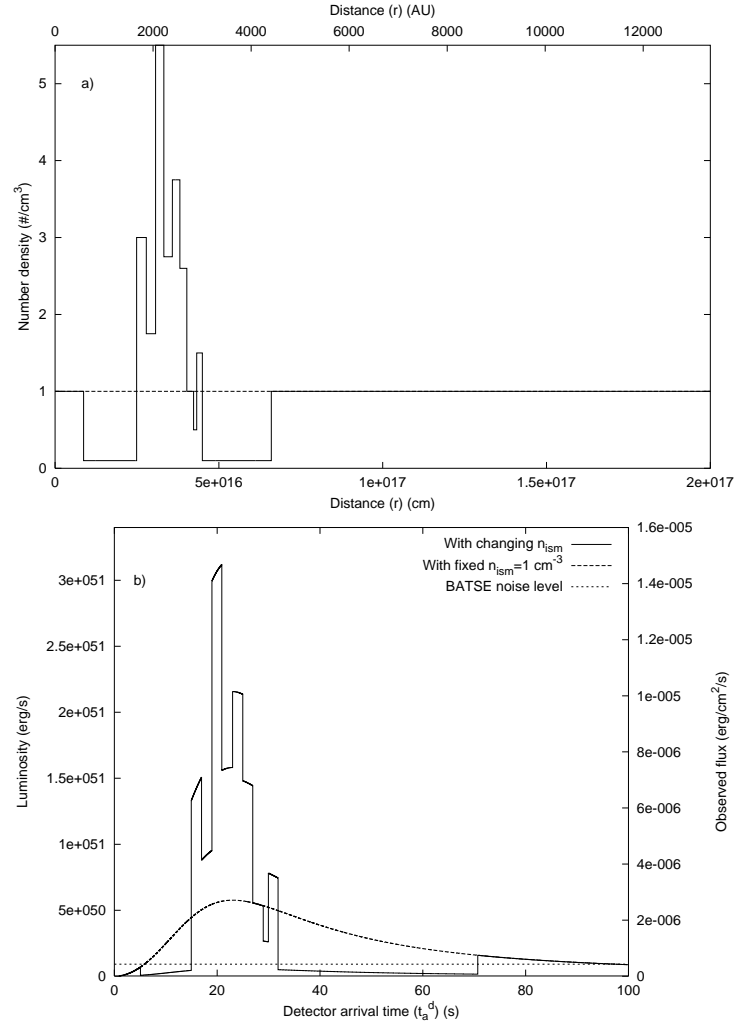}
\end{center}
\caption{a) The density contrast of the ISM cloud profile introduced in order to fit the observation of the burst of GRB~991216. The dashed line indicates the average uniform density $ n = 1 cm^{-3} $. b) Flux computed in the collision of the ABM pulse with an ISM cloud with the density profile given in a). The dashed line indicates the emission from an uniform ISM with $ n = 1 cm^{-3} $. The dotted line indicates the BATSE noise level. Note that 20 seconds in arrival time do corresponds to $\sim 5.0\times 10^{16}$ cm in laboratory frame! Details are given in Table~1 of \cite{ruffini_rbcfx02}. Compare and contrast these theoretical curves with the actual data reported in Fig.~\ref{ruffini_grb991216}.}
\label{ruffini_denscont}
\end{figure}

We have just stressed how the analysis of the sole afterglow of GRB~991216 
data, obtained by BATSE and the Chandra and RXTE satellites, has allowed 
to fix the only two free parameters of the EMBH theory. As a first byproduct of this analysis we can conclude, at variance with results in \cite{ruffini_ha00,ruffini_pk01} pointing to a sharply collimated beamed emission in GRB~991216, that no evidence of beaming is found as a consequence of the perfect agreement between the observed slope of the afterglow and the theoretical value obtained within the EMBH model.

We can now proceed to acquire the predictions of the EMBH theory with reference to two 
fundamental quantities and their role within the GRBs structure: the P-GRB 
and the peak emission of the afterglow.

It soon appeared clear that the :

\begin{enumerate}
\item the so called ``long bursts'' observed by BATSE are 
actually not  bursts at all. Once the proper space-time greed is given 
it is immediately clear that the long bursts are 
generated at distances of $~4\times 10^{16}$ cm from the EMBH. The long 
burst coincides with the Extended-Afterglow-Peak-Emission, which we will 
call E-APE, of the afterglow: they were interpreted as bursts only due to 
the high threshold of the BATSE detectors see Fig.~\ref{ruffini_fit_1}.
\item The time variability observed in simply due to the density 
inhomogeneities intrinsic in an interstellar cloud, as the ABM pulse 
impact on it \cite{ruffini_lett5}, also see Fig.~\ref{ruffini_denscont}. 
\item The previous two conclusions are based on the simplified pure radial 
description of the afterglow presented in \cite{ruffini_rbcfx02}. The effects of angular scattering and spreading in the signal has been considered \cite{ruffini_lettS,ruffini_bcrx02}. This more complex approach leads to interesting new results, but does not affect the two above conclusions.  
\end{enumerate}

The hunt of the P-GRB then started. The interest in identifying it is 
mainly because some general relativistic and relativistic quantum field 
theory effects originating in the process of gravitational collapse during 
the formation of the EMBH are, in principle, encoded in the structure of 
the P-GRB. 

\begin{figure}
\begin{center}
\includegraphics[width=12cm,clip]{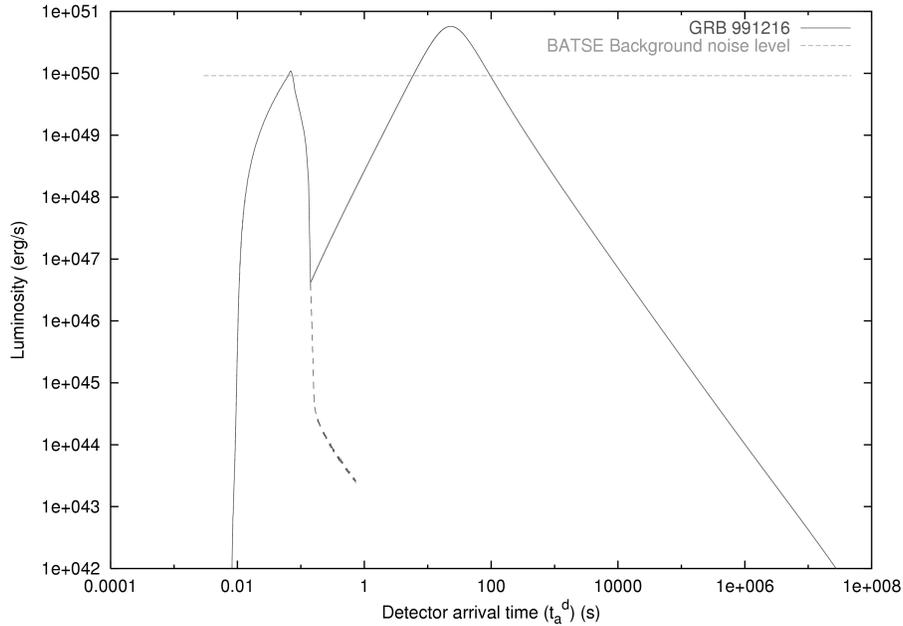}
\end{center}
\caption{Diagram showing the full picture of the model, with both P-GRB 
and E-APE and their relative intensities and time separation. The 
numerical values are presented and tabulated in \cite{ruffini_rbcfx02}.}
\label{ruffini_final}
\end{figure}

Having the only two free parameter of the EMBH theory been fixed, there 
are two fundamental diagrams to be analyzed. The first, 
Fig.~\ref{ruffini_dtab} relates  the precise separation in time 
between the E-APE and the P-GRB, as a function of the amount of baryonic 
matter left over by the gravitational collapse of the progenitor star 
expressed by parameter $B$ and for selected values of the $E_{dya}$. The second relates the intensities of the 
P-GRB to the E-APE, in units of the $E_{dya}$, to the parameter $B$, see Fig.~\ref{ruffini_crossen}. We stress that indeed this last diagram is an universal one, in the adopted variables. From these diagrams 
we can identify with the precision of {\em a few percent} in the intensity and 
with an approximation of {\em a few tenth of milliseconds} the P-GRB with the 
``precursor'' in the BATSE data, (see Fig.~\ref{ruffini_grb991216} and 
Fig.~\ref{ruffini_final}).

Before concluding we would like to stress one final important 
consequence and prediction of the EMBH model:
\begin{enumerate}
\item the most general GRB is composed of the P-GRB and and the afterglow 
(see Fig.~\ref{ruffini_final}). The relative intensity of the P-GRB and the E-APE 
is a function of the $B$ parameter.
\item for $B < 3.5\times10^{-5} $ the energy of the P-GRB is larger then 
the one of the E-APE and the energy of the dyadosphere is mainly emitted in 
what have been called the ``short bursts''. Their afterglow have been 
systematically lower then the BATSE threshold.
\item for $B > 3.5\times10^{-5} $ the energy of the E-APE predominates and 
the energy of the dyadosphere is mainly carried by the ABM pulse and 
emitted in the afterglow. The corresponding E-APE have been improperly 
called ``long bursts''.
\end{enumerate}
It is interesting that this classification also explains at once the 
recently found conclusion that the distribution of short GRBs and long 
GRBs have essentially the same characteristic peak luminosity \cite{ruffini_s01} and the fact that the short 
bursts are systematically harder then long bursts, see \cite{ruffini_rbcfx02} for details. 

Similarly the application of the RSTT and IBS paradigm has naturally conduced as to the concept of induced gravitational collapse \cite{ruffini_lett3}, in order to explain the observed iron lines emission in the late phases of the afterglow of GRB~991216 and analogous effects in three additional GRBs showing clear correlation with supernovae. This is the topic of fortcoming publications \cite{ruffini_rbcx02}.

The understanding of the role of P-GRB and E-APE in GRB~991216, the fact 
that both their absolute and relative intensities have been predicted 
within a few percent accuracy as well as that their arrival time has been 
computed with the precision of a few milliseconds, see \cite{ruffini_rbcfx02}, and 
Figs.~\ref{ruffini_fit_1},\ref{ruffini_dtab},\ref{ruffini_final}, can be considered one of 
the major success of the EMBH theory. 

New space missions have to be conceived to explore, on additional GRB 
sources, the theoretical predictions in the first $10^2$ seconds of 
Fig.~\ref{ruffini_final}. This region has been left vastly unexplored by the BATSE 
data due to the high threshold. Current missions are exploring with great accuracy mainly the later phases of the afterglow. These observations as well as the ones of short bursts, which in the EMBH theory are P-GRB emissions, are indeed crucial, since indeed we re-hiterate: {\em all general relativistic and relativistic quantum field 
theory effects originating in the process of gravitational collapse during 
the formation of the EMBH are, in principle, encoded in the structure of 
the P-GRB} \cite{ruffini_crv02,ruffini_rv02,ruffini_rcvx02}. 

The long lasting debate, started in Princeton in 1971, of how an EMBH is formed has also by now been clarified in \cite{ruffini_r02mg9}. The needed charge segregation process occurs in the magnetosphere of a rotating magnetized star. The charged collapsing core, surrounded by an oppositely charged remnant, approach the EMBH final stages in $\sim 30$ seconds for a $10M_\odot$ progenitor star. The leading process of discharge of the EMBH is due to the vacuum polarization process in view of their very short time scale $10^{-19}$ seconds, see \cite{ruffini_prx02}.

The EMBH theory can now be applied to all other GRB sources.

\end{document}